# Vapor Cloud Delayed-DPPM Modulation Technique for nonlinear Optoacoustic Communication

Muntasir Mahmud, Mohamed Younis, Md Shafiqul Islam, Fow-Sen Choa and Gary Carter
Department of Computer Science and Electrical Engineering,
University of Maryland Baltimore County
Baltimore, Maryland, USA
mmahmud1, younis, mdislam1, choa, carter@umbc.edu

***Abstract*— The optoacoustic process can solve the longstanding challenge of wireless information transmission from an airborne unit to an underwater node (UWN). The nonlinear optoacoustic signal generated by proper laser parameters can propagate long distances in water. However, forming such a signal requires a high-power laser, and the buildup of a vapor cloud precludes the subsequent acoustic signal generation. Therefore, pursuing the traditional on-off keying (OOK) modulation technique will limit the data rate and power efficiency. In this paper, we analyze different modulation techniques and propose a vapor cloud delayed-differential pulse position modulation (VCD-DPPM) technique to improve the data rate and achieve high power efficiency for a single stationary laser transmitter. The symbol rate of VCD-DPPM is approximately 6.9 times and 1.69 times higher than OOK in our text communication simulation using a laser repetition rate of 10 kHz and 40 Hz, respectively. Furthermore, VCD-DPPM is 137% more power efficient than the OOK technique for both cases. We have generated different acoustic signal levels in laboratory conditions and simulated the bit error rate (BER) for different depths and positions of the UWN, while considering ambient underwater noises. Our results indicate that VCD-DPPM enables efficient data transmission.**

***Keywords:* *Air-water communication, Optoacoustic, Modulation, DPPM, Power efficiency, Bit error rate.***

## I. Introduction

Current and future applications are to leverage increased connectivity to enable pervasive data collection and processing. Analogous to the Internet-of-things (IoT), the notion of internet of underwater things (IoUT) captures these characteristics in the realm of underwater applications. The vision of researchers is to establish an internet of everything where underwater nodes become members of a bigger network spanning land, air and seas [1]. Yet, despite the major advances in communication technologies, wireless transmission from air to submerged nodes in water has been challenging. Little progress has been made on transmitting data from a terrestrial IoT connected device to an IoUT connected device because no single signal type can operate well in this cross-medium scenario. High-frequency radio signals are mostly used for terrestrial communications but attenuate very rapidly after entering the water. One approach to circumvent this difficulty is to use very low frequency (VLF) (3 to 30 kHz) and extremely low frequency (ELF) (3 to 30 Hz) bands because they are less vulnerable to attenuation in the water environment. However, broadcasting and capturing such low frequency electromagnetic waves requires huge and cumbersome equipment in both the airborne or land-based transmitter and the underwater receiver. Although visible light communication (VLC) can be effective for shorter underwater ranges, a visual light beams scatters quickly and cannot support the longer communication range [2]. An acoustic signal is the prime choice for underwater communication due to its less attenuation in water; yet it cannot penetrate the air-water interface due to high acoustic impedance mismatch. Therefore the conventional method uses a surface floating node, e.g., buoy or boat, that is equipped with radio and submerged acoustic transceivers to communicate over air and through underwater, respectively. However, deploying such a gateway (surface node) has many significant shortcomings, especially logistical constraints and security risks.

The optoacoustic effect is a viable option for tackling this air-water cross-medium communication challenge [3]. Moreover, optoacoustic signals can be used to localize the UWNs remotely from the air without any gateway or anchor node [4]. The optoacoustic energy conversion mechanism can be subdivided into linear and nonlinear domains based on the energy density and irradiance imparted to the transparent medium like water. The nonlinear optoacoustic process is more efficient for generating the acoustic signal that is more suitable for long distance communication. Many naval applications require an acoustic source level (SL) greater or equal to 180 dB, which the nonlinear process can generate. However, at various times during laser transmission with a high repetition rate, the formation of vapor clouds prevents the generation of subsequent acoustic signals [5], which limits the achievable data transmission rate. Moreover, a high power laser is required for the optical breakdown in the nonlinear process. Therefore traditional modulation techniques are not suitable for efficient wireless data transmission from the air to underwater.

The most common modulation technique in optical communication is OOK and is known for its bandwidth efficiency and high bit rate. In OOK, "0" bit is represented by zero intensity and "1" bit is represented by positive intensity. However, OOK cannot achieve a high data rate in nonlinear optoacoustic communication due to the formation of the vapor cloud during multiple laser beam transmission ("1" bit) in a short time while using a high repetition rate laser. On the other hand, pulse position modulation (PPM) is more power efficient than OOK [6] and can transmit data with a high repetition rate laser without forming the vapor cloud. Nevertheless, PPM is not bandwidth efficient where each $M$ bits are sent over $L = 2^M$ time slots. Thus, the data rate decreases with $M$'s increase despite using a high repetition laser. Moreover, both OOK and PPM require accurate time synchronization between transmitter and receiver, which is extremely difficult to achieve in underwater environments. Although differential pulse position modulation (DPPM) and its improved version IDPPM are more bandwidth efficient than PPM, there can be a vapor cloud buildup in such encoding techniques and limit the achievable data rate.

In this paper, a modification of DPPM referred to as VCD-DPPM is proposed to potentially overcome the challenges in achieving a higher data rate and power efficiency in wireless communication from air to water using a single stationary laser. VCD-DPPM is similar to pulse time modulation where each symbol has a single "1" bit; yet enough delay between the "1" bits in two symbols is provisioned to prevent the vapor cloud formation. Therefore we can transmit data with a high repetition rate laser where the time for each slot is minimized to boost the data rate. We have experimentally validated VCD-DPPM in laboratory setup and simulated the symbol rate for text communication with American Standard Code for Information Interchange (ASCII) characters. The results show VCD-DPPM has approximately 1.69 times and rises up to 6.9 times higher symbol rate than OOK using a laser repetition rate of 40 Hz and 10 kHz, respectively, and is 137% more power efficient than the OOK. To assess the effect of the underwater communication range, we have also simulated the BER in different depths and positions using our experimental results, while considering the underwater ambient noise.

The paper is organized as follows. Section II covers related work. The optoacoustic signal generation and control challenges are analyzed in Section III. Section IV describes our proposed VCD-DPPM modulation and encoding technique for nonlinear optoacoustic communication. Section V reports the performance results. The paper is concluded in Section VI.

## II. Related work

For long underwater reach, the nonlinear optoacoustic signal generation is more effective than the linear mechanism, where a greater source level is achieved; yet the characteristics of the produced underwater acoustic signals are hard to control, which complicates the design of modulation schemes. Overall, the generated acoustic signal has a broadband spectrum up to a few MHz [7]. Blackmon and Antonelli [5] demonstrated a means of deterministically controlling the spacing between the frequency components in the acoustic signal spectrum by varying the laser pulse repetition rate. Therefore they proposed frequency shift keying (FSK), frequency hopped direct sequence spread spectrum (FH-DSSS) technique where a number of laser pulses at the same repetition rate within a given time period represents a symbol [8]. However, the overall acoustic spectrum remained broadband and vulnerable to noise and travel time error. Moreover, several laser pulses are needed to represent a symbol that increases power consumption. Ji et al. [9] proposed a technique to minimize the laser energy requirement using floating low-cost passive relays, which absorb laser pulse energy to generate the acoustic signal by thermal expansion-contraction. These relays could be released by the underwater receiver on the water surface. However, this passive relay needs to be released by every receiver during every data receiving period, which is impractical. Moreover, the thermal absorption generated linear optoacoustic process has a lower communication range than the nonlinear process.

The Naval Research Laboratory (NRL) demonstrated plasma shape control, enabling the acoustic pulse duration control method [10]. As a result, NRL's variable pulse duration method significantly improves with better spectral separation and is more robust to noise and propagation delay errors. Recently we have proposed an Optical Focusing-based Adaptive Modulation (OFAM) technique where the generated acoustic source level can be controlled by varying the laser focusing geometry [11]. OFAM has the potential to increase the bit rate by generating multiple SLs. However, all the above methods utilizing nonlinear optoacoustic effects are vulnerable to vapor cloud formation if a proper encoding technique is not used. Thus, long distance underwater wireless communication from the air using nonlinear optoacoustic signals requires proper modulation and encoding techniques to prevent vapor cloud generation and increase power efficiency. To the best of our knowledge, no prior work on modulation and encoding opts to improve the bit rate and achieve high power efficiency for nonlinear optoacoustic communication.

## III. Nonlinear Optoacoustic Signal Generation and Control Challenges

The optoacoustic process can generate underwater acoustic signals from a remote, aerial location using a high intensity laser transmitter. In the nonlinear optoacoustic process, the physical state of the medium changes during energy transfer, and optical breakdown occurs. This optical breakdown leads to plasma generation at locations where the breakdown threshold is surpassed. The breakdown threshold is dependent on the laser parameters. The irradiance threshold values are in the order of $10^{11}$ W/cm$^2$ for a few nanosecond pulses and rise up to $10^{13}$ W/cm$^2$ for 100 femtosecond pulses to generate optical breakdown in water [12]. The plasma formation during the optical breakdown is associated with the breakdown shockwave and the subsequent cavitation bubble expansion-collapse shockwaves, which generate the acoustic signal.

The requirement of a high-energy laser that needs to be concentrated in a small spot area underwater is one of the challenges in optoacoustic communication. NRL investigated a remote acoustic source generation method where a laser pulse propagates many meters underwater, then quickly converges to a high intensity that exceeds the breakdown threshold at a predetermined location [13]. A combination of group velocity dispersion (GVD), which provides longitudinal compression, and nonlinear self-focusing (NSF), which provides transverse compression, is used to achieve controlled underwater compression of these laser pulses. The relative strength of the GVD and NSF effects of a properly designed laser pulse is such that it can travel hundreds of meters through the air relatively unchanged, then quickly compress upon entering the water.

Another difficulty is the broadband spectrum of the generated acoustic signal, which makes frequency-based modulation quite challenging and limits the communication range. The energy spectral density (ESD) of the generated acoustic signal is dependent on the volume and shape of the plasma, where more elongated plasma volumes produce longer-duration acoustic pulses containing more energy at low frequencies. T.G. Jones et al. proposed a method of generating meter-scale elongated plasma with ESD centered near 1 kHz [14]. However, it requires two laser sources where the first laser generates underwater optical filament, and the second laser pulse is used to heat the filament to create an extended superheated plasma. In order to generate elongated plasma using a single laser transmitter, higher energy is required with a proper focusing angle [10]. Moreover, a higher pulse energy laser can generate higher acoustic SL enabling a greater communication range with the cost of higher power consumption. Thus, a suitable encoding technique is necessary to minimize power consumption.

Vapor cloud generation in the vicinity of the laser transmission in water is a major challenge for nonlinear optoacoustic communication. Blackmon et al. [5] first observed no acoustic transient generation in some laser pulse intervals during higher repetition laser transmission. He concluded that it was most likely caused by the buildup and formation of a vapor cloud in the vicinity of the laser beam transmission area. The vapor cloud effect can be mitigated by dynamic laser transmission or using multiple lasers to fire laser pulses in different positions in the water. However, a single stationary laser transmitter requires a proper encoding technique, considering the delay necessary to prevent vapor cloud formation. In addition to the vapor cloud, there is the possibility of nonlinear interaction of successive laser pulses when the laser pulse repetition rate is higher than 1000 Hz [5]. Therefore, approximately 1 ms of relaxation time is required between two subsequent laser pulses.

## IV. VCD– DPPM Design

### A. Vapor Cloud Delay

Unlike radio, acoustic, and visual light, using optoacoustic communication involves two distinct signal types, optical (laser beam) and acoustic. Therefore, the laser beam needs to be modulated such that the resulting acoustic signals can be demodulated accurately to retrieve the data. Such a communication technique is unique and traditional schemes cannot achieve a higher bit rate. In the previous section, we have discussed the vapor cloud formation and high laser pulse energy requirement. Based on that, we can conclude that the number of "on" chips ("1" bit) should be minimized during data transmission to reduce the power consumption and introduce an unavoidable vapor cloud delay between two "on" chips to enable communication with high repetition rate lasers. This delay means "off" chips ("0" bit), which is to prevent vapor cloud formation. As a result, the bit rate increases because the chip duration ($T_c$) is lower for a higher repetition laser.

The vapor cloud delay can be determined by transmitting continuous "1" bits and increasing the laser repetition rate until the acoustic signal starts missing in some laser pulse intervals. Consequently, vapor cloud delay is calculated from the maximum repetition ($R_{max}$) rate where all the acoustic signals are generated for continuous "1" bits transmission and the vapor cloud delay ($T_v$) is equal to $\frac{1}{R_{max}}$. In [5], optoacoustic signal generation was demonstrated using 200, 500, and 1000 Hz laser repetition rates where several acoustic signals were missed even at 200 Hz. Thus, the vapor cloud delay is greater than 5 ms for their laboratory conditions. However, the exact vapor cloud delay cannot be calculated from their experimental results because they have not shown the maximum repetition rate where all the acoustic signals are generated.

### B. Analysis of Different Modulation Techniques

Despite being the most popular modulation technique in optical communications, OOK has the potential for several consecutive "1" bit transmissions, causing the laser repetition rate to be limited to $R_{max}$. On the other hand, PPM can transmit data with a comparatively higher repetition rate laser because, in $L$-PPM, only one "1" bit and $L$-1 "0" bits are transmitted. Thus, there can be a delay between two "1" bits of two PPM symbols during transmission. However, there is a possibility of two consecutive "1" bits scenario, e.g., if two consecutive symbols are "0001" and "1000" for $M$ = 2. Therefore, the maximum repetition rate of the PPM technique ($R_{max,PPM}$) is equal to the highest repetition rate at which two consecutive "1" bits can be transmitted without creating the vapor cloud effect. $R_{max,PPM}$ is also limited because the delay between two consecutive "1" bits decreases as the laser repetition rate grows, which increases the potential of vapor cloud formation and nonlinear interaction of successive laser pulses. Moreover, PPM is not a bandwidth efficient technique where the number of bits per symbol increases with $M$.

The DPPM technique is used to improve the bandwidth efficiency of PPM, where all the "0" bits are deleted following the "1" bit from the corresponding PPM symbol. However, DPPM also has the potential of several consecutive "1" bits transmission; thus, the maximum repetition rate of the DPPM is the same as OOK. IDPPM is derived from DPPM by adding one extra zero before DPPM symbols [15]. Thus, the IDPPM technique's worst-case scenario is repetitive "01" bits transmission, e.g., the "010101…" bit sequence, but there is at least one "0" bit between two "1" bits. Therefore the maximum repetition rate for the IDPPM technique is twice of $R_{max}$. The bit rate of OOK, PPM, DPPM, and IDPPM is calculated by,

$$B_{OOK} = R_L \tag{1}$$

$$B_{PPM} = \frac{M \times R_L}{2^M} \tag{2}$$

$$B_{DPPM} = \frac{2 \times M \times R_L}{2^M + 1} \tag{3}$$

$$B_{IDPPM} = \frac{2 \times M \times R_L}{2^M + 3} \tag{4}$$

Here, $R_L$ is the laser repetition rate. The maximum value of $R_L$ to prevent vapor cloud effect for OOK and DPPM is $R_{max}$, for IDPPM is $2 \times R_{max}$ and for PPM is $R_{max,PPM}$. Nevertheless, none of these modulation techniques can achieve high bit rate due to their limited laser repetition rate.

### C. VCD-DPPM Technique

VCD-DPPM opts to boost the achievable data rate by using a higher repetition rate laser where $T_c$ is minimized. In VCD-DPPM, symbols are derived from DPPM by adding extra "0" bits before the DPPM symbols, as shown in Fig. 1. The main idea of VCD-DPPM is to include sufficient vapor cloud delay using "0" bits so that the data can be transmitted with a high repetition rate laser without creating the vapor cloud effect. The required number of "0" bits ($N_0$) for vapor cloud delay is equal to $\frac{R_L}{R_{max}} - 1$. $N_0$ can also be represented by $\frac{T_v}{T_c} - 1$. In both cases, the value of $N_0$ needs to be rounded up if it is a fraction.

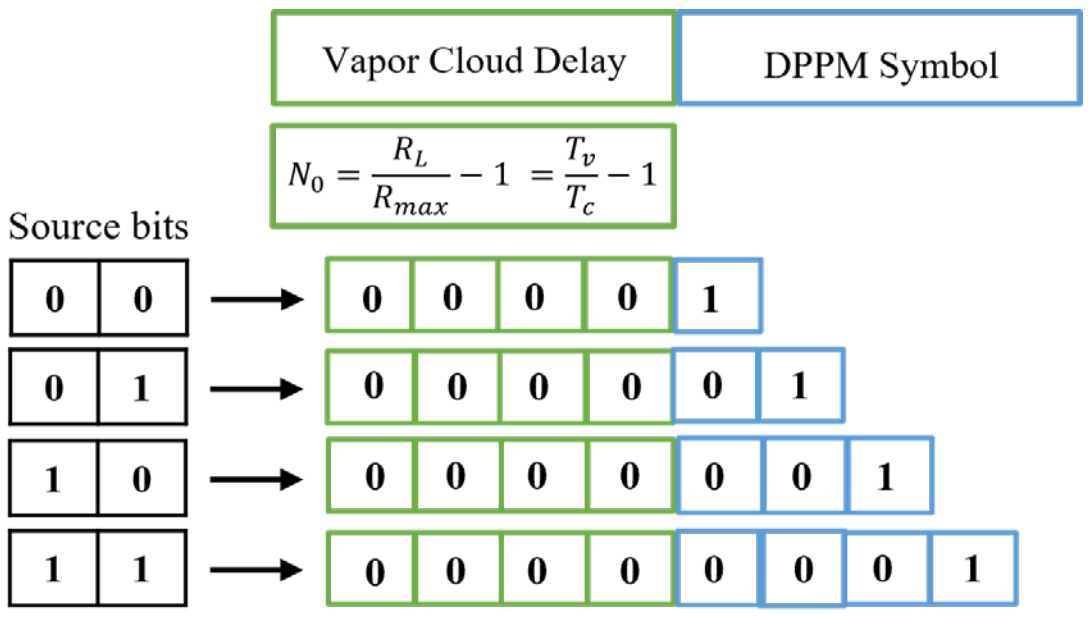

Fig. 1: Illustrating the VCD-DPPM symbol structure with an example for M = 2. Four "0" bits are required to mitigate the effect of vapor cloud.

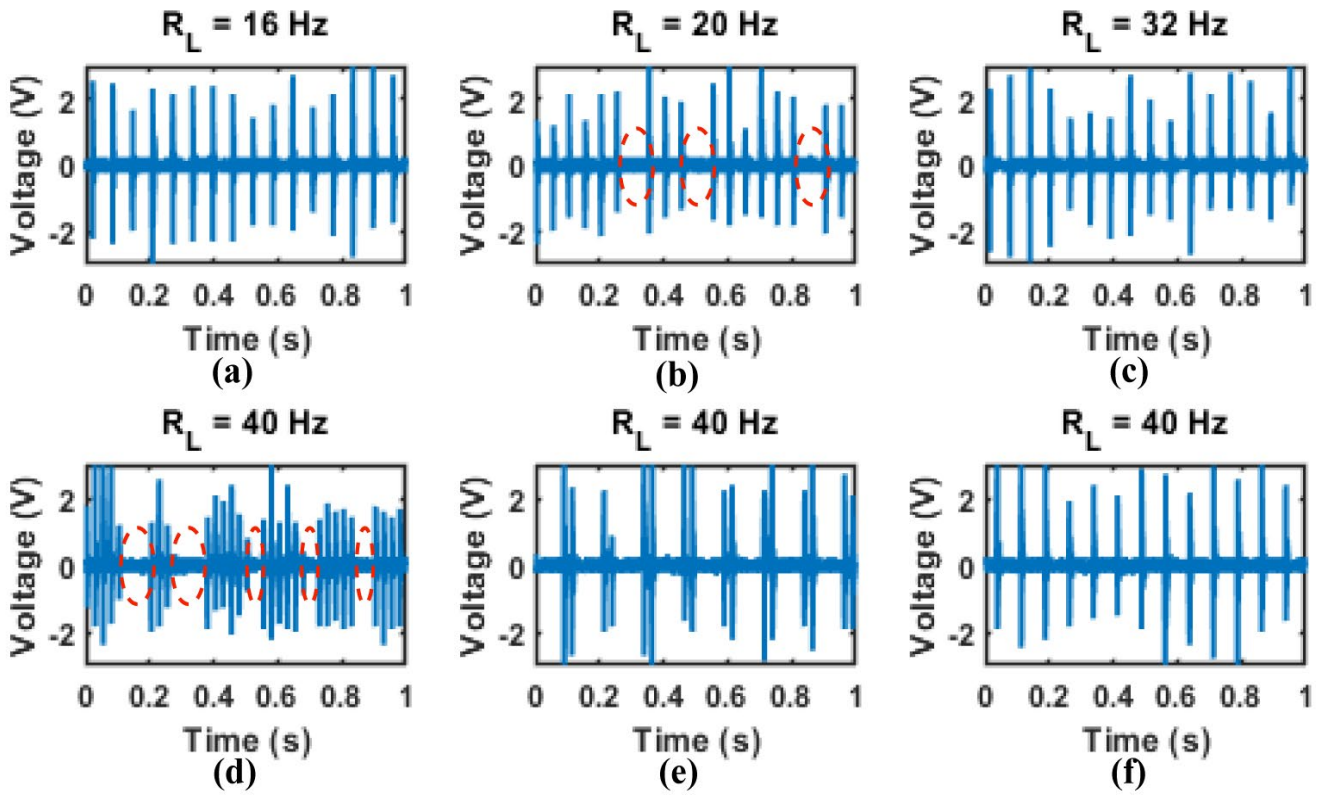

Fig. 2: Experimental optoacoustic signal generation by varying laser repetition rate up to 40 Hz: (a) maximum repetition rate is 16 Hz to transmit continuous "1" bit where no vapor cloud effect is observed. The missing acoustic signals are shown with dotted circles in (b) 20 Hz, and (d) 40 Hz, due to vapor cloud formation. The continuous acoustic signal is generated using (c) "01" repetitive bit sequence in 32 Hz, (e) "11000" repetitive bit sequence in 40 Hz, and (f) "001" repetitive bit sequence in 40 Hz.

Therefore, the data rate of VCD-DPPM increases with the laser repetition rate and decreases as the vapor cloud delay diminishes. The bit rate of VCD-DPPM is calculated by,

$$B_{VCD-DPPM} = \frac{2 \times M \times R_L}{2N_0 + 2^M + 1} \qquad (5)$$

VCD-DPPM symbols have unequal durations and do not require symbol synchronization, similar to DPPM and IDPPM. On the receiver side, the UWN will receive only one acoustic signal for each symbol and demodulate based on the total delay before receiving an acoustic signal. In Fig. 1, an example of VCD-DPPM symbol mapping is shown for $M = 2$ source bits where four "0" bits ($N_0 = 4$) are needed before DPPM symbols. Although the VCD-DPPM technique has more average bit length than OOK, it can produce a higher bit rate because it is not limited by the laser repetition rate. The data rate of VCD-DPPM can be higher than other techniques even at a lower repetition rate if we consider probability-based symbol mapping, where the VCD-DPPM symbols with the lowest number of bits are mapped with symbols with the highest probability of occurrence. If the probability of "0" and "1" bits occurrence is the same in OOK symbols, the power efficiency of the VCD-DPPM with respect to OOK given by [16],

$$P_{VCD-DPPM/OOK} = \left(1 + \frac{M-2}{M}\right) \times 100\% \qquad (6)$$

Eq. (6) indicates that VCD-DPPM is more power efficient than OOK for higher values of $M$. Eq. (6) is also true for PPM, DPPM, and IDPPM because they also have only one "on" chip ("1" bits) in each symbol. Therefore all the techniques have the same power efficiency with respect to OOK.

## V. Performance Analysis

### A. Experimental Results

A laboratory test was conducted to demonstrate nonlinear optoacoustic signal generation using different modulation techniques. We have used a Geoscience Laser Altimeter System (GLAS) Q-switched Nd:YAG laser emitting 6 ns pulses at 1064 nm with a repetition rate up to 40 Hz. The laser pulses are focused on the water with a 7.5 cm lens. A TC4041 hydrophone with a frequency range of 15 Hz to 480 kHz is used to record the underwater acoustic signal generated by the laser. We have transmitted continuous "1" bit and found $R_{max}$ to be 16 Hz, where no acoustic signal is missing and beyond 16 Hz, the vapor cloud effect degrates acoustic signal generation. Thus, the vapor cloud delay is 62.5 ms in our laboratory condition. The maximum repetition rate of IDPPM is 32 Hz, twice than $R_{max}$ as expected, shown in Fig. 2 (c) for the repetitive '01' bit sequence. In Fig 2(e), we have transmitted the repetitive "11000" bit sequence, which is the worst bit sequence scenario in PPM techniques at 40 Hz, and found no acoustic signal missing. Therefore, PPM can be used with a higher repetition rate than OOK, DPPM, and IDPPM. The number of vapor cloud delayed "0" bits required for the VCD-DPPM technique at 40 Hz is two, and we have validated the VCD-DPPM symbol transmission with a repetitive "001" bit sequence shown in Fig 2(f). We have observed in our experiments that vapor cloud buildup after some consecutive "1" bits transmission when a higher repetition rate is used. For example, at a 40 Hz repetition rate, at least 3 consecutive "1" bit transmission forms a strong enough vapor cloud, precluding subsequent acoustic signal generation. Hence the vapor cloud effect can be avoided at 40 Hz if we include vapor cloud delay (using "0" bits) before three consecutive "1" bit transmissions.

### B. Data Rate and Power Efficiency

The bit rate of different modulation techniques for optoacoustic communication is shown in Fig. 3(a) using Eq. (1)-(6). The maximum repetition rate of OOK, PPM, DPPM, and IDPPM is 16 Hz, 40 Hz, 16 Hz, and 32 Hz, respectively, as determined from the experiments. The bit rate of VCD-DPPM at 40 Hz is only highest for $M = 3$. However, the bit rate of VCD-DPPM increases with laser repetition rate, and it has the highest bit rate for all the values of $M$ at 300 Hz. We can observe that the bit rate of VCD-DPPM decreases for higher $M$ values because the average bit length increases with $M$. Fig. 3(b) compares the power efficiency of the different modulation techniques to OOK. The power efficiency increases with $M$ because the number of "1" bits grows for larger $M$ values in OOK but remains the same for all other techniques.

To illustrate the performance of our proposed VCD-DPPM technique, we have done a text communication simulation using the ASCII characters in a novel written by William Morris entitled: "The Well at the World's End". The novel is provided in a text file of size 1.186 MB taken from the Gutenberg Project [17]. The probabilities for the characters shown in Table 1 were obtained by computing the occurrences of the characters in the text file (not all of them are shown in the table). The symbol mapping is done based on the probability to achieve a higher data rate, as shown in Table 1, where the symbols with the lowest number of bits are assigned to the characters with the highest probability. Fig.4 (a) shows the symbol rate for a laser repetition rate up to 40 Hz based on Table 1's mapping. Even for the large $M = 7$ value, VCD-DPPM achieves the highest symbol rate. The symbol rate of IDPPM ($S_{IDPPM}$) is close to VCD-DPPM because we are considering only up to 40 Hz laser.

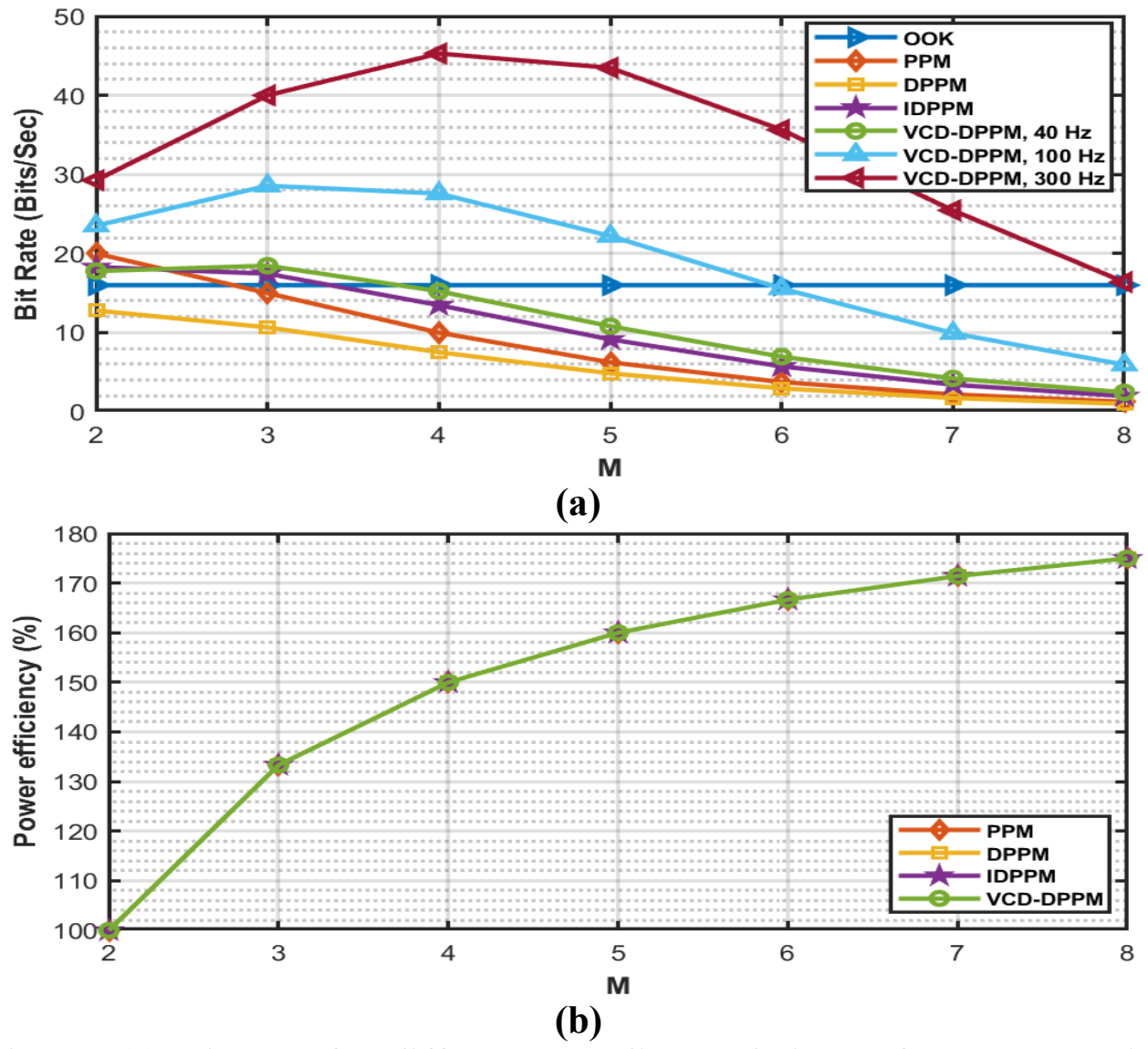

Fig. 3: (a) Bit rate for different encoding techniques for optoacoustic communication, (b) Power efficiency relative to OOK for different encoding techniques.

TABLE 1: Frequency of some ASCII characters in a novel and their corresponding symbol mapping for different modulation techniques.

| Character | Probability | OOK | PPM | DPPM | IDPPM | VCD-DPPM |
|---|---|---|---|---|---|---|
| space | 0.18095053 | 0000000 | 10000000000000000000000000000000000000000000000000000000000000000000000000000000000000 | 1 | 01 | 001 |
| e | 0.09747676 | 0000001 | 01000000000000000000000000000000000000000000000000000000000000000000000000000000000000 | 01 | 001 | 0001 |
| t | 0.07361019 | 0000010 | 00100000000000000000000000000000000000000000000000000000000000000000000000000000000000 | 001 | 0001 | 00001 |
| a | 0.06544206 | 0000011 | 00010000000000000000000000000000000000000000000000000000000000000000000000000000000000 | 0001 | 00001 | 000001 |
| h | 0.0633835 | 0000100 | 00001000000000000000000000000000000000000000000000000000000000000000000000000000000000 | 00001 | 000001 | 0000001 |
| o | 0.05759156 | 0000101 | 00000100000000000000000000000000000000000000000000000000000000000000000000000000000000 | 000001 | 0000001 | 00000001 |
| n | 0.04862421 | 0000110 | 00000010000000000000000000000000000000000000000000000000000000000000000000000000000000 | 0000001 | 00000001 | 000000001 |
| s | 0.0423335 | 0000111 | 00000001000000000000000000000000000000000000000000000000000000000000000000000000000000 | 00000001 | 000000001 | 0000000001 |
| i | 0.03971266 | 0001000 | 00000000100000000000000000000000000000000000000000000000000000000000000000000000000000 | 000000001 | 0000000001 | 00000000001 |
| d | 0.03935898 | 0001001 | 00000000010000000000000000000000000000000000000000000000000000000000000000000000000000 | 0000000001 | 00000000001 | 000000000001 |
| . | . | . | . | . | . | . |
| . | . | . | . | . | . | . |
| . | . | . | . | . | . | . |

In Fig. 4(b), we have simulated up to 10 kHz repetition rate to show the increase of the symbol rate of VCD-DPPM ($S_{VCD-DPPM}$) in comparison to other techniques. We can observe that, $S_{VCD-DPPM}$ is below 4 symbols/second at 40 Hz but goes up to approximately 16 symbols/second in Fig. 4(b). PPM has the worst symbol rate ($S_{PPM}$) among the considered modulation techniques in Fig. 4(a) because of the increased number of bits in each symbol. However, all the bits are "0" bits in PPM symbols except one bit; these "0" bits can be utilized as a vapor cloud delay, enabling communication with a higher repetition rate laser. In Fig. 4(b), we have assumed that PPM can use up to a 500 Hz repetition rate based on the results of the experiments shown in [5], where three consecutive "1" bits are required to create the vapor cloud effect at 500 Hz. Therefore, PPM performs better than OOK, DPPM, and IDPPM at higher repletion rates in Fig. 4(b). $S_{VCD-DPPM}$ increases with the laser repetition rate and achieves approximately 6.9, 2.7, 8.2, and 4.6 times higher symbol rate at 10 kHz than OOK, PPM, DPPM, and IDPPM, respectively. We can observe that $S_{VCD-DPPM}$ has a steep increase up to 1000 Hz repetition rate but does not increase as much for higher repetition rates due to 62.5 ms vapor cloud delay consideration. Therefore, the symbol rate of VCD-DPPM will increase more if the vapor cloud delay can be decreased using different laser parameters. Fig. 4(c) illustrates the power efficiency where all the techniques achieve 137% more power efficiency than OOK in this simulation.

We have also experimentally validated VCD-DPPM using Table 1's mapping and successfully transmitted "Hello underwater world!" from air to underwater using the 40 Hz laser repetition rate, as shown in Fig. 5. At the beginning of the data transmission, four control bits- "1001" are transmitted to initialize the receiver and calculate the threshold by averaging the peak to peak voltage generated by the two acoustic signals. Then, based on the threshold calculated by the receiver, "1" bits are detected and decoded based on the "0" bits (no acoustic signal) received before a single "1" bit (acoustic signal).

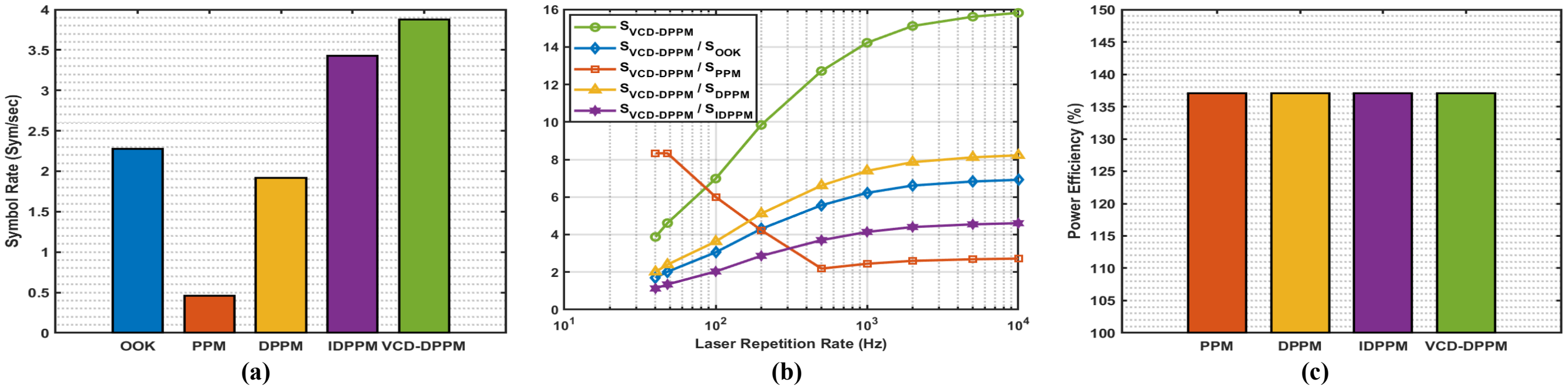

Fig. 4: (a) Symbol rate of the different modulation techniques up to 40 Hz laser repetition rate based on the mapping in Table 1, (b) simulation results of the VCD-DPPM technique for higher laser repetation rate and symbol rate increase respect to other techniques, and (c) power efficiency increase respect to OOK.

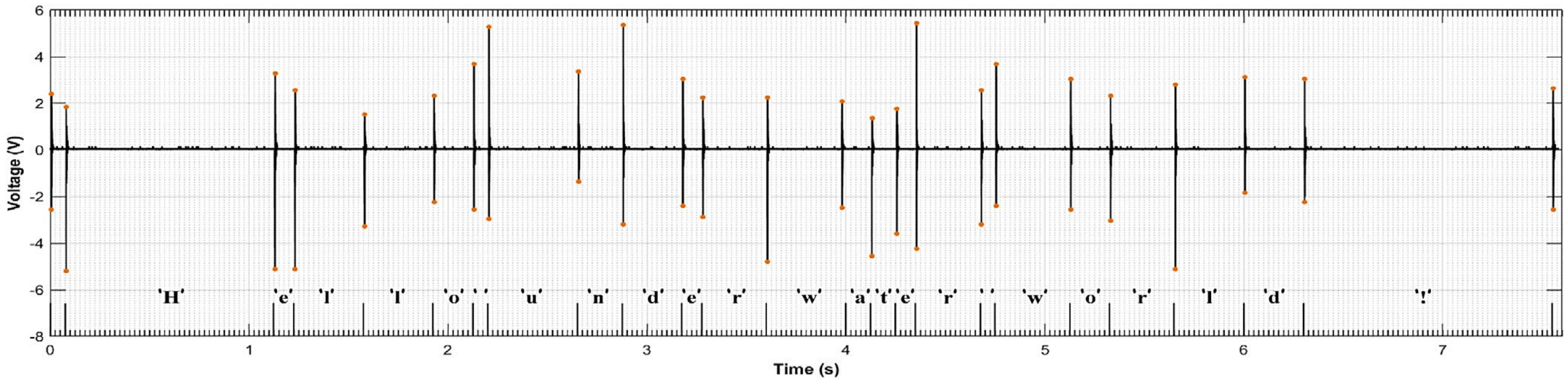

Fig. 5: Experimental demonstration of the VCD-DPPM technique for optoacoustic communication using "Hello Underwater world!" text transmission. At the beginning control bits -"1001" are transmitted to initialize and calculate the threshold value for demodulation.

### C. *BER Simulation*

We have simulated the BER in Matlab for different depths and positions using our experimental results to evaluate the underwater communication range. Different SL values of the acoustic signal at $0^0$, $45^0$, and $90^0$ relative to the laser beam incident are generated by varying the laser pulse energy in the laboratory, as shown in Fig. 6(a) - (d). The SL value decreases as the acoustic signal propagates underwater due to the transmission loss (TL). Therefore, the sound intensity level (SIL) received by the UWN at a distance (D) from the acoustic source plasma can be determined by subtracting the TL from the SL values. The TL calculation is done similar to [4]. First, 64 control bits are transmitted to the UWN to calculate the threshold, and then $10^5$ data bits are sent using Table 1. Fig. 6(e) shows the BER at 500 m underwater using our experimental SL values while assuming the underwater ambient noise of [18]. We can observe that VCD-DPPM can achieve acceptable BER even at a 500 m distance underwater for 50 mJ and 60 mJ laser pulse energy. The BER of VCD-DPPM is better than OOK in all three positions because fewer "1" bits are in VCD-DPPM symbols than OOK symbols.

## VI. Conclusions

The nonlinear optoacoustic process has the potential to tackle the challenge of long distance wireless communication in the air-water cross-medium scenario. However, the presence of a vapor barrier and the existence of cavitation bubbles due to preceding laser pulse interactions can preclude the subsequent acoustic signal generation, consequently limiting the bit rate. This paper proposes the VCD-DPPM modulation/encoding technique to improve the data rate and achieve high power efficiency. We have validated VCD-DPPM with text communication simulation; the results have shown that VCD-DPPM achieves a higher symbol rate than all the traditional techniques and better power efficiency than OOK. Moreover, the symbol rate of VCD-DPPM increases with the laser repetition rate; meanwhile the laser repetition rate is limited for traditional techniques due to the potential of vapor cloud formation and the nonlinear interaction of successive laser pulses. Simulation using experimental data indicates that VCD-DPPM can achieve great BER even at a 500 m distance underwater. Therefore, VCD-DPPM enables efficient wireless data transmission from the air to long distances underwater.

Acknowledgment: This work is supported by the NSF, contract #0000010465. The authors would like to thank Dr. Anthony Yu from NASA GSFC for providing access to the Q-switch Nd:YAG laser.

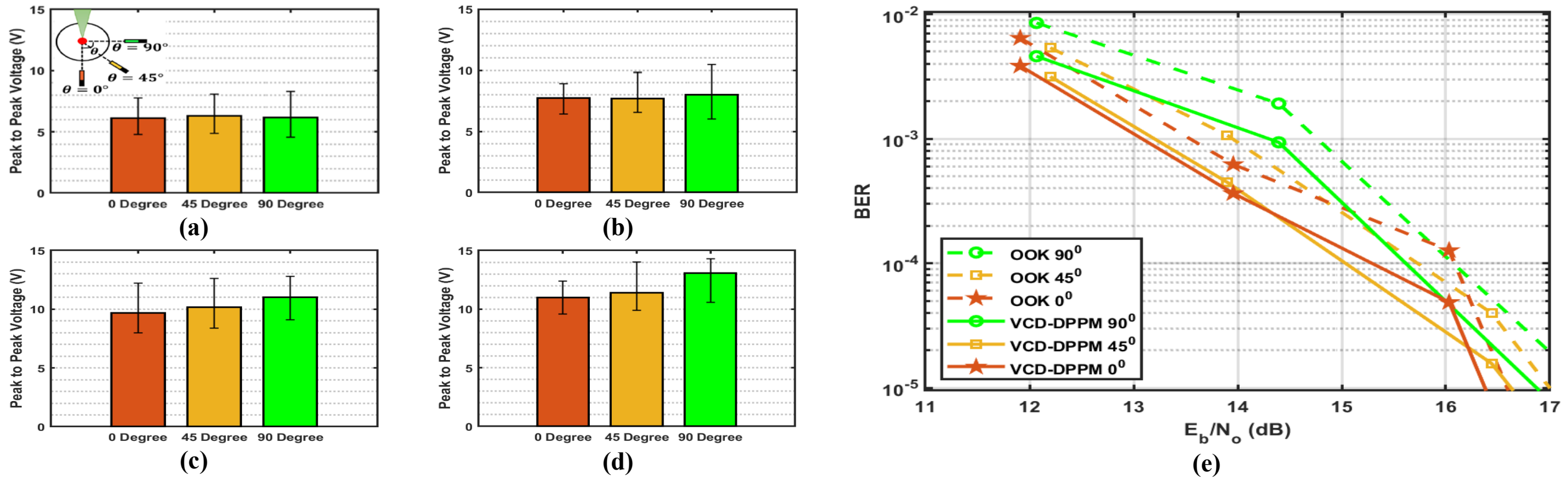

Fig. 6: Peak-peak voltage of experimentally generated acoustic signal using (a) 22 mJ (b) 35 mJ (c) 50 mJ (d) 60 mJ laser pulse energy. Each presented value is the average of at least fifteen measurements with error bar showing the maximum and minimum values, and (e) BER comparison between OOK and VCD-DPPM techniques at 500 m distance from the acoustic source and $0^0$, $45^0$ and $90^0$ receiver positions relative to the laser beam propagation.